\documentclass[referee]{aa}
\usepackage{graphics}

\def\ltapprox{\raise 2pt \hbox {$<$} \kern-1.1em \lower 5pt \hbox {$\approx$}}
\def\ltsim{\raise 2pt \hbox {$<$} \kern-1.1em \lower 4pt \hbox {$\sim$}}
\def\gtsim{\raise 2pt \hbox {$>$} \kern-1.1em \lower 4pt \hbox {$\sim$}}

\begin{document}

\titlerunning{Astrophysical UI powered by GW emission}

\title{Unipolar Inductor Model coupled to GW emission: energy budget and model
application to RX J0806+15 and RX J1914+24}

\author{S.Dall'Osso\inst{1}
\and 
 G.L.Israel\inst{1}
\and 
L.Stella\inst{1} 
}

\offprints{dallosso@mporzio.astro.it}

\institute{
INAF--Osservatorio Astronomico di Roma, via Frascati 33, I--00040 Monteporzio 
Catone
(Roma), Italy; dallosso, gianluca and stella@mporzio.astro.it 
}
\date{}

\abstract{
We further discuss the Unipolar Inductor Model (UIM) coupled to GW emission
(Dall'Osso et al. 2005) and compare it to observed properties of the two
candidate ultrashort period binaries RX J0806+15 and RX J1914+24 .\\
We consider the measured orbital periods, period derivatives and inferred 
X-ray luminosities of these two sources and find constraints on system 
parameters in order for the model to account for them.
We find that these properties point to the two sources being in different 
regimes of the UIM, with the requirement of low magnetic moment primaries 
($\sim 10^{30}$ G cm$^3$) for both.\\
Given this weak magnetization, RX J0806+15 has a sufficiently low luminosity
that it can be interpreted as having
a primary spin almost synchronous to and just slightly slower than the orbital 
motion. Its measured orbital spin-up is only slightly affected by spin-orbit 
coupling and is mostly due to GW emission.\\
RX J1914+24, on the other hand, is too bright in X-rays and has too slow 
an orbital spin-up for the same regime to apply. We suggest that this binary
system may be emitting GWs at a significantly higher rate than implied by its 
measured $\dot{\omega}_o \simeq 6 \times 10^{-17}$ rad s$^{-2}$. The latter is 
explained, in this framework, by the primary spin being slightly faster than 
the orbital motion ($\alpha \leq 1.1$). In this case, the associated spin-orbit
coupling \textit{transfers to} the orbit a significant amount of angular 
momentum, thus partially balancing that lost to GW emission.\\
All expectations can be tested in the near future to confirm the viability of 
the model.
\keywords{}
}

\maketitle

\section{Introduction}
\label{intro}
The nature of the two X-ray sources RX J1914+24 and RX J0806+15 has been long 
debated since their discoveries and the origin of their emission is still quite
unclear. Several models have been proposed so far, none of which  seems able
to explain all of their observed properties without facing major difficulties
(see Cropper et al. 2003 for a review on models and observations).\\
The long-term timing analyses of Strohmayer (2004,2005), Hakala et al. (2003, 
2004) and Israel et al. (2004), the most recent ones based on $\sim$ 10 yrs of 
monitoring, have unambiguously shown that the single periods found in 
the power-spectra of both objects (respectively $\sim$ 9.5 and 5.3 min) are 
steadily decreasing over time. The very nature of such periodicities is 
debated, evidence being presently in favour of its interpretation as the 
orbital period of a binary system. Indeed, no other significant periods are 
found in power spectra which is quite unexpected if, for example, the observed 
periods reflected the spins of accreting compact objects.\\ 
If the observed periods are true orbital modulations, then 
it is extremely difficult to accomodate Roche-lobe filling non-degenerate 
secondary stars orbiting a compact companion in such close binary systems, 
because of the extremely small masses implied. For this reason, 
double-degenerate binary systems represent the most likely option.\\
As orbital periods are observed to decrease, accretion models in a double 
degenerate system are apparently ruled out because (conservative) accretion in 
these cases should lead to a period increase. 
Therefore, although with the above caveats, the double-degenerate scenario with
no mass exchange seems at least more suited than others to explain the timing 
properties of these sources.\\
In this context the Unipolar Inductor Model (UIM), originally proposed for the
Jupiter-Io system (Goldreich \& Lynden-Bell 1969), has been suggested as an 
alternative explanation to the observational properties of the above mentioned 
sources (Wu et al. 2002). In this version of the model two 
white dwarfs form a close binary system with a moderately magnetized primary 
and a secondary that, for practical purposes, is non-magnetic. If the primary 
spin ($\omega_1$) is not perfectly synchronous with the orbital motion 
$\omega_o$ (the secondary is tidally locked to the orbit) an e.m.f. is induced 
across the secondary star, idealized as a perfect conductor crossing the 
primary's magnetic field lines. In the presence of free charges, the e.m.f. 
induces a current flow between the two stars, along the sides of the flux tube 
connecting them. Currents in the atmosphere of the primary cross field lines at
a depth where the electrical conductivity ($\sigma$) becomes isotropical; at 
this point, they close the circuit returning back to the secondary (Goldreich 
and Lynden-Bell 1969, Li, Ferrario and Wickramashige 1998, Wu et al. 2002). \\
Resistive dissipation of currents and the associated heating, that occur 
essentially in the primary atmosphere, cause the observed soft X-ray emission.
The source of the emission is ultimately represented by the mechanical energy 
of the relative motion between the primary spin and the orbit. Therefore, 
dissipation consumes this energy and brings the primary and the orbit 
into synchronous rotation.\\
As shown by Dall'Osso et al. (2005, paper I from here on), perfect 
synchronization is never reached in the UIM applied to ultrashort period DDBs. 
The emission of GW continuously injects energy in the electric circuit when 
$\omega_1 < \omega_o$ and can thus sustain a permanent, slight asynchronism of 
the primary spin. We called this the system's steady-state asynchronism: 
binaries in this regime, \textit{i.e.} with sufficiently short orbital periods,
will have a primary spin just slightly slower than the orbital motion, the 
small asynchronism causing a continuous flow of electric current which is 
ultimately powered by GWs.\\
In this work, we first address and clarify some general issues discussed in 
paper I. Then, we apply our model to the two candidate ultrashort period 
binaries RX J0806+15 and RX J1914+24. We obtain simple relations between a
source luminosity, its \textit{measured} rate of orbital spin-up, component 
masses and primary magnetic moment, that give constraints on system parameters 
for the model to be applicable to both sources.\\
Detailed and systematic calculations and evolutionary implications are beyond 
the scope of this work and will be addressed elsewhere.
\section{The Unipolar Inductor Model: summary}
\label{general}
We summarize here briefly the features of the UIM, referring to Wu et al. 
(2002) and paper I for a more detailed description. The key ingredient of the 
model is that the primary spin ($\omega_1$) differs from the orbital motion
 ($\omega_o$); $\alpha = \omega_1/ \omega_o$ is the asynchronism parameter.\\
In an asynchronous system with orbital separation $a$ the secondary star moves 
across the primary magnetic field lines with a relative velocity $ v = a 
(\omega_o - \omega_1)$. The electric field induced through the secondary is 
{\boldmath$E$} = $\frac{\mbox{{\boldmath$v \times B$}}}{c}$. The associated 
e.m.f. is $\Phi = 2R_2 E$, $R_2$ being the secondary's radius. The induced 
flow of currents is resistively dissipated essentially in the primary 
atmosphere, causing significant local heating and thus powering the soft 
X-ray emission. The Lorentz torque on cross-field currents in the primary 
atmosphere and in the secondary causes spin-orbit coupling, redistributing 
angular momentum between the primary spin and the orbital motion. However,
given the significant GW-emission expected from two white dwarfs with 
ultrashort orbital period, orbital angular momentum is continuously lost from 
the system, which in turn requires spin-orbit coupling to continue: this keeps 
the electric circuit active at all times.\\
Concerning the specific assumptions of this work, we consider binary systems 
hosting two degenerate white dwarfs, with the following mass-radius relation 
(Nauenberg 1972):
\begin{equation}
\label{massradius}
\frac{R}{R_{\odot}} = 0.0112 \left[ \left(\frac{M}{1.433}\right)^{-\frac{2}{3}}
- \left(\frac{M}{1.433}\right)^{\frac{2}{3}}\right]^{\frac{1}{2}}
\end{equation}
where $M$ is expressed in solar masses.\\ 
In paper I we have shown that tidal synchronization of the primary component 
is not expected to be efficient on the orbital evolutionary timescale, while 
it should well be efficient for a lower mass companion, in most cases of 
interest. The system orbital evolution is thus given by (eq. E4 in Appendix E
of Wu et al. 2002):
\begin{equation}
\label{omegadot}
\frac{\dot{\omega}_o}{\omega_o} = \frac{1}{g(\omega_o)}\left(\dot{E}_{gr}
- \frac{W}{1-\alpha}\right)
\end{equation}
where $\dot{E}_{gr} = - (32/5)G/c^5 [q/(1+q)]^2 M^2_1 a^4 \omega^6_o$ is the
energy loss rate through GW emission (Landau \& Lifshitz 1951), the second 
term in parentheses represents the contribution of spin-coupling and W is the 
electric current dissipation rate (source luminosity). The function 
$g(\omega_o) < 0$ (see appendix A for its definition) represents two thirds of 
the orbital (kinetic plus gravitational) energy of the binary system, times a 
coefficient just slightly smaller than unity. The latter accounts for the 
additional tiny amount of orbital angular momentum that is continuously lost 
to keep the secondary into synchronous rotation.\\
GWs give by definition a positive contribution to $\dot{\omega}_o$ while the
contribution of electric coupling ($W > 0$ by definition) depends on the sign 
of $(1-\alpha)$. Given the \textit{measured} orbital spin-up of both sources 
under study, it must be assumed for them that, if $\alpha>1$, then 
$|\dot{E}_{gr}| > |W/(1-\alpha)|$.\\
The quantity W can be expressed as (paper I):
\begin{eqnarray}
\label{W}
W  & = & \frac{\phi^2}{\Re} = \left(\frac{\mu_1}{c}\right)^2 \frac{2 
\overline{\sigma} R^{3/2}_1 R^3_2}{[GM_1(1+q)]^{11/6}} \frac{\omega^{17/3}_o 
(1-\alpha)^2}{(H /\Delta d) \jmath(e)} \nonumber \\
& = & k~\omega^{\frac{17}{3}}_o (1-\alpha)^2
\end{eqnarray}
where $\Re$ is the system effective resistance (see Wu et al. 2002), $\mu_1$ 
the primary's magnetic moment, $\overline{\sigma}$ the height-averaged WD 
atmospheric conductivity, $H$ the atmospheric depth at which currents cross 
magnetic field lines and return back to the secondary and $\Delta d$ is the 
thickness of the arc-like cross section of the current layer at the primary 
atmosphere. The last equality defines the system's constant $k$.\\
Finally, we recall the evolution equation for $\alpha$ (see paper I):
\begin{equation}
\label{alfaevolve}
\frac{\dot{\alpha}}{\alpha} = \frac{W}{\alpha (1-\alpha) I_1 \omega^2_o} -
\frac{\dot{\omega}_o}{\omega_o} ,
\end{equation}
where $\tau_{\alpha} \equiv \alpha/\dot{\alpha}$.
\section{UIM coupled to GW emission: energy budget and the general solution} 
\label{energetics}
In paper I we discussed the conditions that cause the current flow and, thus, 
the soft X-ray emission, never to be quenched by synchronization of the primary
star. In the time-independent, linear approximation, the evolution equation 
for $\alpha$ is solved by a simple exponential function, whose asymptotic value
$\alpha_{\infty} \leq 1$ confirms this expectation.\\
Numerical integration of the full evolution equation confirms that, as the
orbital period changes over time, $\alpha$ does indeed evolve asymptotically 
towards unity, never reaching it exactly. 
We showed that, in the general solution, the value $\alpha^{en}_{\infty}
\leq 1$ is attained, which conserves the total mechanical energy resevoir 
($E_{UIM}$) of the circuit
\begin{equation}
\label{EUIM}
E_{UIM} = (1/2) I_1 (\omega^2_1 -\omega^2_o).
\end{equation}
Therefore, by requiring that the time derivative of $E_{UIM}$ is zero,
%
%
we obtain (paper I)
\begin{equation}
\label{paperone}
 \alpha^{en}_{\infty} (1-\alpha^{en}_{\infty}) = \frac{I_1}{k}
\frac{\dot{\omega}_o/ \omega_o}{\omega^{11/3}_o}.
\end{equation}
This expression for $\alpha^{en}_{\infty}$ is implicit, as it depends on the 
source's actual value of $\dot{\omega}_o$, a function of $\alpha$ 
itself\footnote{The problem has a non-linear nature.}. Despite this, the above 
formula is of great practical use as it relates the model parameter 
$\alpha^{en}_{\infty}$ to measured quantities ($\omega_o$, $\dot{\omega}_o$).
We will return on this in $\S$ \ref{application}.\\
Given the importance of the steady-state solution, it is necessary to specify 
here its physical meaning and definition, starting from a discussion of the 
overall energy budget of the system. 
\subsection{Asynchronism parameter and efficiency of spin-orbit coupling}
\label{efficiency}
As a preliminary step, we summarize the relevant properties of spin-orbit 
coupling in the UIM.
First recall the expression of the Lorentz torque $N_L = W/[\omega_o 
(1-\alpha)]$ (paper I). This same torque $N_L$ acts - with opposite sign -
on both the primary star and the orbit. Therefore, angular momentum is 
conserved by spin-orbit coupling: all that is subtracted from one component is 
transferred to the other one.\\
The rate of work done by $N_L$ on the orbit is
\begin{equation}
\label{Eorb}
\dot{E}^{(orb)}_L = - N_L \omega_o = - \frac{W}{1-\alpha},
\end{equation}
while that done on the primary spin is:
\begin{equation}
\label{espin}
\dot{E}_{spin} = N_L \omega_1 = \frac{\alpha}{1-\alpha} W = -\alpha 
\dot{E}^{(orb)}_L.
\end{equation}
The sum $\dot{E}_{spin} + \dot{E}^{(orb)}_L = -W$, which means that not all of
the energy extracted from one component is transferred to the other one. The 
energy lost to ohmic dissipation is the energetic cost of spin-orbit coupling.
\\
The following picture thus results from the above formulae:
\begin{itemize}
\item When $\alpha>1$ the circuit is powered at the expenses of the primary's 
spin energy. Only a fraction $\alpha^{-1}$ of this energy is transferred to the
orbit, the rest being lost to ohmic dissipation. 
\item When $\alpha <1$ the circuit is powered at the expenses of the orbital 
energy. Now $\alpha$ measures directly the fraction of energy that is 
transferred to the primary spin, while ($1-\alpha$) is that lost to ohmic 
dissipation\footnote{Note that, for $\alpha =0$, $\dot{E}_{spin}=0$ but
$\ddot{E}_{spin} > 0$}.
\end{itemize}
Therefore, the parameter $\alpha$ can be seen as a measure of the efficiency 
with which spin-orbit coupling transfers energy between the orbit and the
primary spin: the more asynchronous a system is, the less efficiently energy 
is transferred, most of it being dissipated as heat.
\subsection{Stationary state and the general solution}
\label{budget}
We can now re-discuss and clarify the energetic argument of paper I, based on 
which we obtained the expression of $\alpha^{en}_{\infty}$.
The steady-state regime is characterized by having a constant electric energy 
reservoir in the circuit, for a given $\omega_o$. The degree of asynchronism
is thus set by the condition that the system uses up just the power
fed to the circuit by GWs and spin-orbit coupling.
It is important to note, in this context, that both ohmic dissipation and the 
energy required to change $\omega_o$ and/or $\omega_1$ must be taken into 
account. Therefore, the condition $\dot{E}_{UIM} =0$ of paper I would not be 
wholly consistent, because only $\dot{\omega}_1$ and $\dot{\omega}_o$ appear 
in $\dot{E}_{UIM}$ and no account is taken of the luminosity W. 
We then re-state the problem of steady-state in the following terms: since the 
circuit dissipates energy at a rate $W$, it must recharge at the same rate as 
the binary evolves, if its energy reservoir is to be conserved. The proper 
steady-state condition is thus
\begin{equation}
\label{stationary}
\dot{E}_{UIM} = \pm W~,
\end{equation}
where the sign on the right-hand side depends on the asynchronism parameter. 
The plus sign holds when $E_{UIM} >0$ and, thus, $\alpha >1$, while the minus 
sign is relevant to the case with $\alpha <1$ ($E_{UIM} <0$).\\
%
%
In paper I (Appendix B) we have shown that the case with $\alpha_{\infty} >1$ 
corresponds to a condition very hardly realized in WD+WD systems, because it 
can hold only for very close systems. So close, indeed, that the lighter 
component would most likely fill its Roche-lobe. Even if it did not, however, 
the condition also requires $\dot{\omega}_o < 0$, which means the system must 
widen; the two components will rapidly be too far apart for the condition to
hold. Therefore, proper steady-state exists in WD+WD systems only with 
$\dot{E}_{UIM} = - W$.\\
Inserting the expressions for $\dot{\omega}_1$ and $W$ in condition 
(\ref{stationary}) we obtain the expression for the steady-state asynchronsim. 
We call it simply $\alpha_{\infty}$, to distinguish it from the (slightly 
different) expression of $\alpha^{en}_{\infty}$ of paper I. 
\begin{equation}
\label{stedisteit}
1 - \alpha_{\infty} = \frac{I_1}{k} \frac{\dot{\omega}_o / \omega_o} 
{\omega^{11/3}_o} = A .
\end{equation}
%
%
%
%
%
Our former assumption $\dot{E}_{UIM} =0$ amounts ultimately to neglecting W 
compared to the energy terms contained in $\dot{E}_{UIM}$. Note that the 
luminosity W is produced by ohmic dissipation, while $I_1 \omega_1 
\dot{\omega}_1 = \dot{E}_{spin} = W \alpha / (1-\alpha)$ becomes much larger 
than that, the more $\alpha$ gets close to unity. Therefore, neglect of $W$ is 
actually justified when $\alpha$ is close to 1, a condition expected to hold in
steady-state. Indeed, the expression of eq. (\ref{stedisteit}) for $1-
\alpha_{\infty}$ gives a correction $\sim$ $1$\% to (\ref{paperone}), for 
typical cases of interest.
\section{Model application: equations of practical use}
\label{application}
Eq. (\ref{stedisteit}) allows to relate directly the model parameter 
$(1-\alpha_{\infty})$ to measured quantities, in view of the application of the
model to real sources. By contrast, a purely numerical treatment would require 
solving the evolution equation for $\alpha$ with given (unknown) initial 
conditions (asynchronism, orbital period, component masses and primary 
magnetic moment), deriving all other quantities as functions of this solution 
and comparing to observations.\\
We have used this latter approach to calculate $(1-\alpha_{\infty})$ in the
numerical solution for ($1-\alpha$): it turns out that, once the system 
reaches the appropriate regime, the two quantities are essentially coincident. 
Their ratio at sufficiently short orbital periods\footnote{typically $\sim 
1000$ s, see the definition of P$_{fast}$ in the next section.} is $\sim 
1.001-1.002$, decreasing with the orbital period down to P $\sim 100$ s or even
less. Therefore, $(1-\alpha_{\infty})$ represents the actual degree of 
asynchronism of systems with sufficiently short orbital period. It can thus 
be used to obtain physical constraints on the two systems directly from 
measured quantities.
%
%
\subsection{The asynchronism evolution timescale}
\label{timescale}
We have shown that the value $\alpha_{\infty}$ (and, accordingly, all 
steady-state quantities) can be determined in full generality from energetic 
considerations and can be related to \textit{measured} quantities and 
fundamental system parameters (as described below). 
The evolution of a system in a short-lived, highly asynchronous regime, can be 
determined only from the integration of the 
evolutionary equation. In this case, indeed, the free-energy reservoir is 
represented by the (unknown) primary's degree of asynchronism.\\
In paper I we have studied the 
evolution equation of $\alpha$ in the linear approximation, with constant 
coefficients. This corresponds to the regime where spin-orbit coupling gives 
just a negligible contribution to the orbital evolution and, at the same time, 
the synchronization timescale $\tau_\alpha$ is $\ll \tau_o$.
A system with an initial arbitrary value of $\alpha$ may not be in the linear 
regime: however, the UIM always tends to bring the primary into almost 
synchronous rotation, so that the linear regime will eventually apply. At this 
point, in order for the condition $\tau_{\alpha} \ll \tau_o$ to hold, $(1-
\alpha_{\infty})$ must be sufficiently small or, equivalently, the orbital 
period sufficiently short. As shown in paper I, this regime is formally reached
for $1-\alpha_{\infty} < 3\times 10^{-2}$ or, equivalently, when the orbital 
period is shorter than the critical period:
\begin{eqnarray}
\label{plim}
P_{\mbox{\tiny{fast}}} & \simeq & 1.2 \times 10^3~\mbox{s} \left(\frac{M_1}
{0.9~M_{\odot}}\right)
^{-\frac{7}{2}} \left(\frac{\mu_1}{2.5 \times 10^{30}}\right)^2
\left(\frac{R_1} {6 \times 10^8}\right)^{\frac{3}{2}}  \nonumber \\
 & & \times  \left(\frac{R_2}{1.7 \times 10^9}\right)^3   
\frac{\left(\frac{I_1}{2.8 \times 10^{50}}\right)^{-1}}{q (1+q)^{\frac{3}{2}} 
\left(\frac{H} {\Delta d}\right)\jmath(e)}.
\end{eqnarray}
where scaling factors are in c.g.s. units.
P$_{\mbox{\tiny{fast}}}$ measures the period at which, in the numerical 
solution, ($1-\alpha$) is very well approximated by ($1-\alpha_{\infty}$).
\\ 
Strictly speaking, then, if the orbital period is below the limiting value 
$P_{\mbox{\tiny{fast}}}$ the linear, time-independent approximation 
applies and $\alpha$ reaches $\alpha_{\infty}$ on a time shorter than $\sim 
0.03$  times the orbital evolutionary timescale.
The probability of catching a system during the short-lived, asynchronous phase
is accordingly very small. Most systems with P $< P_{\mbox{\tiny{fast}}}$ are 
caught in the long-lived steady-state.\\
We finally note that, when $\alpha$ is very close to 1, it can change only by a
factor $\ll \alpha$. Therefore, the timescale $\tau_{\alpha} =\alpha / 
\dot{\alpha}$ can become very long just because of this. However, as $\alpha$ 
passes from 1.1 to 1.001, changing by $\sim$ 10\% only, $(1-\alpha)$ decreases 
by two orders of magnitude and the same holds for $\dot{E}^{(orb)}_L$, while 
the dissipation rate $W$ changes by four orders of magnitude.\\
In this situation, the evolution timescale of $1-\alpha$ provides more direct
information on the evolving properties of the system, because this parameter 
evolves towards zero. We thus define, neglecting the minus sign in the last 
equality:
\begin{equation}
\label{unomenalfa} 
\tau_{(1-\alpha)} = \frac{1-\alpha}{d (1-\alpha)/ d t} =
- \frac{1-\alpha}{\alpha}~\tau_{\alpha} = (1-\alpha_{\infty})~\tau_o
\left[1 - \alpha~\frac{1-\alpha_{\infty}}{1-\alpha}\right]^{-1}
\end{equation}
The difference between this and $\tau_{\alpha}$ is negligible as long as 
$\alpha \gg 1$ and becomes important when $|1-\alpha| < 1$. In particular, for 
$\alpha = \alpha_{\infty}$, $\tau_{\alpha_{\infty}} = \alpha_{\infty}/ 
(1-\alpha_{\infty}) \tau_o$, much longer than $\tau_o$. Indeed, $\alpha$ 
changes only very slightly once it is close to 1.\\
On the other hand, $\tau_{(1-\alpha_{\infty})}= \tau_o$, from which the basic 
meaning of steady-state is recovered clearly: \textit{it is the regime in which
the system's degree of asynchronism evolves on exactly the same timescale of 
the orbital period}.\\
For any $\alpha \neq \alpha_{\infty}$ the corresponding timescale becomes 
quickly much shorter than $\tau_o$, thus hampering the possibility of observing
systems that are too far from steady-state, as stated above.
%
%
%
%
%
\subsection{Relating measurements to system parameters}
\label{relating}
We can now summarize the formulae that allow a direct comparison with the 
measured timing properties and inferred luminosities of the two sources.\\
The key parameter ($1-\alpha_{\infty}$) depends only on the measured quantities
 $\dot{\omega}_o$ and $\omega_o$ (with no assumption on the relative strength 
of GW and spin-orbit coupling), on the primary moment of inertia $I_1$ and on 
the unknown parameter $k$, the value of which is primarily determined by 
$\mu_1$. Therefore, based on measurements of $\omega_o, \dot{\omega}_o$ and
$W$, we can construct the quantity $\alpha_{\infty}$ as a function of $k$ and 
use it to estimate $\mu_1$. In particular, given the definition of 
$\alpha_\infty$ and $W$ (eq. \ref{stedisteit} and \ref{W}, respectively), we 
have
%
%
%
%
%
\begin{equation}
\label{luminosity}
W = I_1 \omega_o \dot{\omega}_o~\frac{(1-\alpha)^2}{1-\alpha_{\infty}}.
\end{equation}
The source steady-state luminosity can thus be written as:
\begin{equation}
\label{luminsteady}
W_{\infty} = I_1 \dot{\omega}_o \omega_o (1-\alpha_{\infty})
\end{equation}
%
%
The equation for the orbital evolution (\ref{omegadot}) provides a further 
relation between the three measured quantities, component masses and degree of 
asynchronism.
%
%
We re-write it here in a form that makes the dependence on measured quantitites
and unknown parameters explicit:
\begin{equation}
\label{useful}
\dot{E}_{gr} - g(\omega_o) (\dot{\omega}_o / \omega_o) =  \frac{W} 
{(1-\alpha)} \nonumber
\end{equation}
that becomes, inserting the appropriate expressions for $\dot{E}_{gr}$ and 
$g(\omega_o)$ (cfr. $\S$ \ref{general}), and neglecting the small correction
to $g(\omega_o)$ due to tidal locking of the secondary:
\begin{equation}
\label{extended}
\frac{32}{5}\frac{G^{7/3}}{c^5}~\omega^{10/3}_o X^2  -\frac{1}{3}~
G^{2/3} \frac{\dot{\omega}_o}{\omega^{1/3}_o} X + \frac{W}{1-\alpha} = 0~,
\end{equation}
where $X \equiv M^{5/3}_1 q/(1+q)^{1/3} = {\cal{M}}^{5/3}$, $\cal{M}$ being the
system's chirp mass.\\
With all above equations we can now turn to each of the two systems and assess
which region of the UIM parameter space applies to each of them.
\section{RX J0806+15}
\label{rxj08}
\begin{figure}[h]
\includegraphics{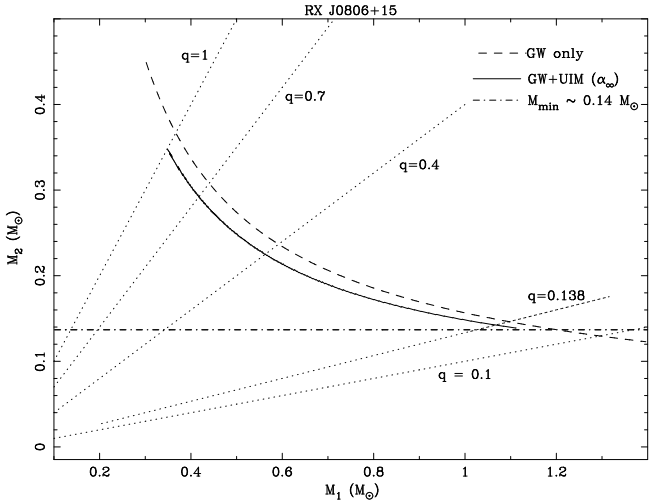}
\vspace{6.7cm}
\caption{$M_2$ vs. $M_1$ plot based on the measured timing properties of 
RX J0806+15. The dashed curve is the 
locus expected if orbital decay is driven by GW alone, with no spin-orbit 
coupling. The solid line describes the locus expected if the 
system is in a steady-state, with $(1-\alpha) = (1-\alpha_{\infty}) \simeq 8 
\times 10^{-3}$. The horizontal dot-dashed line represents the minimum mass 
for a degenerate secondary not to fill its Roche-lobe at an orbital period of 
321.5 s. Dotted lines are the loci of fixed mass ratio.} 
\label{fig3}
\end{figure}

\begin{figure}[h]
\includegraphics{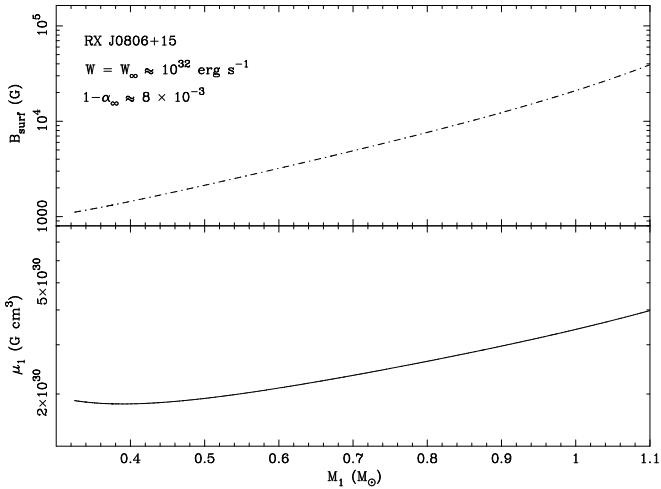}
\vspace{7.0cm}
\caption{The value of the primary magnetic moment $\mu_1$, and the 
corresponding surface B-field, as a function of the primary mass M$_1$, for
$(1-\alpha) = (1-\alpha_{\infty}) = 8\times 10^{-3}$.}
\label{fig2}
\end{figure}

Israel et al. (2003) measured an on-phase X-ray luminosity (in the range 
0.1-2.5 keV) $L_X = 8 \times 10^{31} (d/200~\mbox{pc})^2$ erg s$^{-1}$ for 
this source. These authors suggested that the bolometric luminosity might even
be dominated by the (unseen) value of the UV flux, and reach values 5-6 times
higher (\textit{i.e.} L$_{\mbox{\tiny{bol}}} \sim 5 \times 10^{32}$ erg 
s$^{-1}~d^2_{200}$). The optical flux is only $\sim$ 15\% pulsed, indicating 
that most of it might not be associated to the unipolar inductor mechanism 
(possibly the cooling luminosity of the white dwarf plays a role). Given these 
uncertainties and, mainly, the uncertainty in the distance to the source, we 
assume here a luminosity $W\simeq 10^{32} (d/200~\mbox{pc})^2$ erg s$^{-1}$. 
In $\S$ \ref{change} we consider the effect of assuming a larger source 
luminosity for RX J0806+15 (and RX J1914+24 as well), and show that conclusions
are affected only weakly.\\
Israel et al. (2004) and Strohmayer (2005) obtained independent, phase-coherent
timing solutions for the orbital period of this source over a $\sim$ 10 yrs
baseline, that are fully consistent within the errors. The solution reported
by Israel et al. (2004) is $\dot{P} = -3.67(1) \times 10^{-11}$ and $P = 
321.53033(2)$. These give $\dot{\omega}_o \simeq 2.23 \times 
10^{-15} $ rad s$^{-2}$ and $\omega_o \simeq 0.0195 $ rad s$^{-1}$.\\
In Fig. \ref{fig3} (see caption for further details), the dashed line 
represents the locus of points in the $M_2$ vs. $M_1$ plane, for which the 
measured $\omega_o$ and $\dot{\omega}_o$ are consistent with being due to GW 
emission only, \textit{i.e.} if spin-orbit coupling was absent ($\alpha =1$). 
This corresponds to a chirp mass ${\cal{M}} \simeq$ 0.3 M$_{\odot}$. \\
Inserting the measured quantities in eq. (\ref{luminosity}) and assuming a 
reference value\footnote{Possible values of $I_1$ range in a limited interval: 
$I_1= 3.6 \times 10^{50}$g cm$^2$ for $M = 0.2~M_{\odot}$, $I_1 = 3.75 \times 
10^{50}$ g cm$^2$ for $M = 0.5~M_{\odot}$, $I_1 = 3.2 \times 10^{50}$ g cm$^2$ 
for $M = 0.8~M_{\odot}$, $I_1=2.4 \times 10^{50}$ g cm$^2$ if $M_1=1~M_{\odot}$
and $I_1\simeq 10^{50}$ g cm$^2$ for $M_1 = 1.3~M_{\odot}$.} of 
$I_1 = 3 \times 10^{50}$ g cm$^2$, we obtain:
\begin{equation}
\label{j08}
\frac{(1-\alpha)^2} {1-\alpha_{\infty}} \simeq \frac{10^{32}d^2_{200}}
{1.3 \times 10^{34}} \simeq 8 \times 10^{-3} d^2_{200}.
\end{equation}
In principle, the source may be in any regime, but its short orbital period 
strongly suggests that it is beyond the limiting period 
$P_{\mbox{\tiny{fast}}}$ of eq. (\ref{plim}). Therefore, the most likely and
natural scenario is that it has already settled into steady-state and $\alpha =
\alpha_{\infty}$. \\
In this case, we see from eq. (\ref{j08}) that $(1-\alpha_{\infty}) \simeq 8 
\times 10^{-3}$.
Once UIM and spin-orbit coupling are introduced, the locus of allowed points in
the M$_2$ vs. M$_1$ plane is somewhat sensitive to the exact value of $\alpha$:
the solid curve of Fig. \ref{fig3} was obtained for $\alpha = \alpha_{\infty} 
= 0.992$. $M_1$ must be smaller than 1.1 $M_{\odot}$ in order 
for the secondary not to fill its Roche lobe, thus avoiding mass transfer. 
Note that, if M$_1 > 1.2$ M$_{\odot}$, $I_1 \leq 10^{50}$ g cm$^2$ and 
assuming $W=W_{\infty}$ would imply $1-\alpha_{\infty} > 8 \times 10^{-3}$.
In this case, the solid curve would be shifted downwards, crossing the 
minimum-mass horizontal line at M$_1 < 1$ M$_{\odot}$. In turn this would rule
out the high M$_1$ values based on which the curve was obtained. Therefore,
we conclude that, if RX J0806+15 is interpreted as being in the UIM 
steady-state, $M \leq 1$ M$_{\odot}$ is necessarily required.\\
From $(1-\alpha_{\infty}) = 8\times 10^{-3}$  we obtain, by use of 
eq. (\ref{W}), $k \simeq 7.7 \times 10^{45}$ (c.g.s.): from this, 
component masses and primary magnetic moment can be constrained (see Fig. 
\ref{fig3}). Indeed, since $k = \hat{k}(\mu_1, M_1, q; \overline{\sigma})$ (see
its definition in eq. \ref{W}), we have a further constraint on it, because 
$M_1$ and $q$ must lie along the solid curve of Fig. \ref{fig3}. Given the 
value of $\overline{\sigma}$, $\mu_1$ is obtained for each point along the 
solid curve.\\
According to Wu et al. (2002), and references therein, the electrical 
conductivity of a white dwarf atmosphere with a temperature $\sim 10^5$ K has a
value $\sim 10^{13} \div 10^{14}$ (e.s.u.). We take $\overline{\sigma} = 
3\times 10^{13}$ (e.s.u.), so that the whole range above is within a factor 3.
\\
The values of $\mu_1$ obtained in this way, and the corresponding field at the 
primary's surface, are plotted in Fig. \ref{fig2}, from which $\mu_1 \sim$ a 
few $\times 10^{30}$ G cm$^3$ results, somewhat sensitive to the primary mass.
\\
We note further that the solid curve of Fig. \ref{fig3} is not strictly a locus
of constant chirp mass. However, it is nearly so: $X \simeq (3.4 \div 4.5) 
\times 10^{54}$ g$^{5/3}$ along it, which implies ${\cal{M}} \simeq (0.26 \div 
0.31)$ M$_{\odot}$. More importantly, $\dot{E}_{\mbox{\tiny{gr}}} \simeq (1.1 
\div 1.9) \times 10^{35}$ erg s$^{-1}$ and, since $W/(1-\alpha_{\infty}) = 
\dot{E}^{(orb)}_L \simeq 1.25 \times 10^{34}$ erg s$^{-1}$, we have 
$\dot{E}_{\mbox{\tiny{gr}}} \simeq (9\div 15)~\dot{E}^{(orb)}_L $. Therefore, 
orbital spin-up is essentially driven by GW alone; indeed, the dashed and solid
 curves are very close on the M$_2$ vs. M$_1$ plane.
Summarizing, we have shown that RX J0806+15 has observational properties that
can be well interpreted in the UIM framework, assuming it is in steady-state.
This requires the primary to be moderately magnetic, with $\mu_1 \sim 10^{30}$ 
G cm$^3$ and with an almost synchronous spin, slightly slower than the 
orbital motion (the difference being less than $\sim 1$\%). 
The small degree of asynchronism is consistent with the fact that no modulation
of pulse arrival times at the beat period $\omega_b = \omega_o - \omega_1$ is 
observed, since its small amplitude would likely be below the intrinsic timing
noise of the source (cfr. Barros et al. 2004). 
The expected value of the magnetic moment can also be tested by future 
observations, for example through studies of polarized emission, at optical 
and/or radio wavelenghts (Willes \& Wu 2004).
\section{RX J1914+24}
\label{rxj19}
The luminosity of this source has been subject to many revisions over the last
few years. Wu et al. (2002) refer to earlier ASCA measurements that, for a 
distance of 200-500 pc, corresponded to a luminosity in the range ($4\times 
10^{33} \div 2.5 \times 10^{34}$) erg s$^{-1}$. Ramsay et al. (2005), based on
more recent XMM-Newton observations and a standard blackbody fit to the X-ray
spectrum, derived an X-ray luminosity of $\simeq 10^{35} d^2_{kpc}$ erg 
s$^{-1}$, where $d_{kpc}$ is the distance in kpc. The larger distance of 
$\sim$ 1 kpc was based on a work by Steeghs et al. (2006). Still more recent 
XMM-Newton observtions have been analyzed by Ramsay et al. (2006): these 
authors find that an optically thin, thermal emission spectrum, with an edge at
0.83 keV (attributed to O VIII) gives a significantly better fit to the data 
than the previously used blackbody model. Compared to a blackbody, the 
optically thin thermal plasma model clearly leads to a source bolometric 
luminosity much lower than previously derived, L$_{\mbox{\tiny{bol}}} \simeq 
10^{33}$ d$^2_{kpc}$ erg s$^{-1}$. Ramsay et al. (2006) also note that the 
determination of a 1 kpc distance is not free of uncertainties and that a much
smaller distance ($\sim 200$ pc) is still possible: based on the latter, the 
luminosity would become as small as $\sim 3 \times 10^{31}$ erg s$^{-1}$. \\
Given these large discrepancies, interpretation of this source's properties is
clearly ambiguous and quite dependent on which assumptions are made. We refer 
here to the more recent assessment by Ramsay et al. (2006) of a luminosity 
$L = 10^{33}$ erg s$^{-1}$ for a 1 kpc distance, but want to stress mainly the 
general characteristics of our procedure. The numerical estimates that we 
obtain are, at this stage, necessarily illustrative and, in $\S$ \ref{change}, 
we show how derived quantities change when different luminosities are assumed.
\\
Ramsay et al. (2006) also find possible evidence, at least in a few 
observations, of two secondary peaks in power spectra. These are very close to
($\Delta \nu \simeq 5\times 10^{-5}$ Hz) and symmetrically distributed around 
the strongest peak at $\sim 1.76 \times 10^{-3}$ Hz, the latter usually 
interpreted as the orbital period of the source. We will briefly comment on 
these possible features at the end of $\S$ \ref{lifetime}.\\
Ramsay et al. (2005), confirmed by Ramsay et al. (2006), also give the most up 
to date measurement of the source's rate of orbital evolution, $\dot{\mbox{P}} 
-3.2(10) \times 10^{-12}$, that converts to $\dot{\omega}_o \simeq 6.2 \times 
10^{-17}$ rad s$^{-2}$ for the orbital period of $569$ s, or $\omega_o \simeq 
0.011$ rad s$^{-1}$. We warn that these parameters were not obtained, like in 
the case of RX J0806+15, through a phase-coherent analysis, therefore they may 
be subject to larger uncertainties and future revision.\\ 
Application of the scheme used for RX J0806+15 to this source is not as 
straightforward, since its properties indicate that it is not in a steady-state
and, likely, not even in the regime where spin-orbit coupling is negligible in 
the orbital evolution.\\
First of all, its inferred luminosity seems inconsistent with being associated 
to steady-state as, with the measured values of $\omega_o$ and 
$\dot{\omega}_o$, eq. (\ref{luminosity}) gives\footnote{again assuming $I_1
= 3 \times 10^{50}$ g cm$^2$}
\begin{eqnarray}
\label{j19luminosity}
\frac{(1-\alpha)^2}{1-\alpha_{\infty}} & = & \frac{10^{33}}{2
\times 10^{32}}~\mbox{d}^2_{kpc} \nonumber \\ 
 & \simeq & 5 \mbox{d}^2_{kpc}~,
\end{eqnarray}
which has important implications.
First of all, given $\omega_o$ and $\dot{\omega}_o$, the system steady-state 
luminosity should be $ \ll 2 \times 10^{32}$ erg s$^{-1}$ (eq. 
\ref{luminsteady}), hardly consistent with measurements. Indeed, even if a 
large value of $(1-\alpha_{\infty}) \simeq 10^{-1}$ was assumed, $W \simeq 2 
\times 10^{31}$ erg s$^{-1}$, less than the smallest possible luminosity 
reported by Ramsay et al. (2006). A relatively high ratio between the actual 
asynchronism parameter and its steady-state value appears unavoidable. Indeed, 
from the above equation:
\begin{equation}
\label{j19}
|1-\alpha| \simeq  2.2 (1-\alpha_{\infty})^{1/2}
\end{equation}
For example, if this system had the same $k$ and $I_1$ as RX J0806+15, it would
have a 2.5 times smaller value of $(1-\alpha_{\infty}) \simeq 3.3 
\times 10^{-3}$, from which $|1-\alpha| \simeq 0.13$. 
%
\subsection{The case for $\alpha>1$}
\label{thecase}
The low rate of orbital shrinking implied by the measured $\dot{\omega}_o 
\sim 6.2 \times 10^{-17}$ s$^{-2}$ (Ramsay et al. 2005) suggests a 
very low mass system, with $M_1 \leq 0.3~M_{\odot}$; in this case the secondary
would not fill its Roche-lobe only for $q\geq 0.26$ (see Fig. \ref{fig5} and
Nelemans 2004 for a discussion of this point). In the UIM framework, this 
result gives important hints as to the primary spin.\\
If, as shown above, the system is far from steady-state and has $\alpha <1$, 
the locus of allowed masses in the M$_2$ vs. M$_1$ plane is shifted downwards 
with respect to the dashed line of Fig. \ref{fig5} by a significant amount. 
This would exacerbate the problem of the very low component masses. 
In particular because, $\alpha$ being quite different from 1, the downward 
shift of the curve in the M$_2$ vs. M$_1$ plane would accordingly be large.\\
This argument can be made more cogent by using the measured value of 
$\dot{\omega}_o > 0$; a strong constraint on $|1-\alpha|$ results, because the 
high inferred luminosity $W\sim 10^{33}$ erg s$^{-1}$ would imply a high value 
of $\dot{E}^{(orb)}_L$. If $\alpha<1$, this latter term sums to the GW torque: 
the resulting orbital evolution would thus be faster than if it were driven by 
GW alone. In fact, the smallest possible value of $\dot{E}^{(orb)}_L$ for 
$\alpha <1$ is obtained with $\alpha = 0$, from which $\dot{E}^{(orb)}_L = 
10^{33}$ erg s$^{-1}$. \\
Therefore, if $\alpha <1$, an absolute minimum to the system's rate of orbital 
shrinking would result from the above minimum rate of orbital energy extraction
operated by spin-orbit coupling. This minimum $\dot{\omega}_o$ turns out to be
very close to the measured one, so that it would imply unplausibly small 
component masses in order for $\dot{E}_{\mbox{\tiny{gr}}}$ to be sufficiently 
small. This can be shown by comparing the corresponding chirp masses.
The dashed curve in Fig. \ref{fig5} corresponds to a value of $X \simeq 1.07 
\times 10^{54}$ g$^{5/3}$, or a chirp mass ${\cal{M}} \simeq 0.13$ M$_{\odot}$.
On the other hand, an equivalent GW luminosity of $10^{33}$ erg s$^{-1}$ would 
correspond to $X \simeq 8.5 \times 10^{53}$ g$^{5/3}$ or an equivalent chirp 
mass ${\cal{M}}\simeq 0.12$ M$_{\odot}$.\\ 
Therefore, the slow measured rate of orbital evolution of this source 
essentially rules out the case $\alpha <1$ in the UIM framework discussed here:
indeed, no point in the M$_2$ vs. M$_1$ plane can be found for $\alpha<1$ if a
non-Roche-lobe-filling secondary is required (as it is in the UIM). We note 
that the absolute minimum is obtained assuming $\alpha =0$, a very unlikely 
condition. Summarizing, Fig. \ref{fig5} and the above argument strongly lead to
consider the case where the primary spin is faster than the orbital motion. 
This would indeed offer a great advantage in interpreting this source: in the 
case $\alpha>1$, spin-orbit coupling has an opposite sign with respect to the 
GW torque. The two mechanisms have opposite effects on the orbital evolution 
and a slow measured $\dot{\omega}_o$ can result from the two opposite torques 
partially cancelling each other.\\
We want to stress further the above arguments, as there appears to be some 
confusion in the literature.
Marsh \& Nelemans (2005) have recently claimed that, in order for the UIM to 
account for the properties of RX J1914+24, $\alpha \geq 10^2$ would be 
required, implying the primary is spinning nearly at break-up.\\
These authors arrive at their conclusion first deriving an estimate of the
GW luminosity from the measured $\omega_o$ and $\dot{\omega}_o$ and then
requiring that $W/(1-\alpha)$ be less than the estimated 
$\dot{E}_{\mbox{\tiny{gr}}}$. Given the high source luminosity (they assume 
$10^{35}$ erg s$^{-1}$) and small $\dot{\omega}_o$, a large value of 
$|1-\alpha|$ is required in order for that condition to hold. Note that, 
assuming W$\simeq 10^{33}$ erg s$^{-1}$, their argument gives $|1-\alpha|>1$, 
not so extreme indeed.
Most importantly, however, this argument relies on the assumption that the 
measured timing parameters of this source give a reliable estimate of its 
actual GW luminosity.
This assumption holds only if spin-orbit coupling is negligible; we have shown 
above that this needs not be the case and, most likely, is not the case in this
source. The argument by Marsh \& Nelemans might thus be reversed and used as 
evidence that a non-negligible spin-orbit coupling is necessary in applications
of the UIM to this source.\\
Ramsay et al. (2006) suggest that the much smaller luminosity of $10^{33}$ erg 
s$^{-1}$ they derive would weaken the objection by Marsh \& Nelemans, since it 
would require an asynchronism $(1-\alpha) \sim 10^{-3}$ only (or even less). It
is easy to see that this is not true, because with W$=10^{33}$ erg s$^{-1}$ and
$(1-\alpha) \leq 10^{-3}$ one would obtain $\dot{E}^{(orb)}_L = W/(1-\alpha) 
\geq 10^{36}$ erg s$^{-1}$, which is greater than any plausible GW emission 
from this system, given its measured $\dot{\omega}_o$. Rather, $|1-\alpha|$ 
cannot be too small without making the model inconsistent with the observed 
slow orbital spin-up.\\
The conclusion of this section can only be that, in a self-consistently 
constructed UIM framework, the GW luminosity of this source has to be larger 
than implied by $\dot{\omega}_o$, which implies $\alpha >1$, otherwise the 
model hardly applies. 
\subsection{Constraining the asynchronous system}
\label{lifetime}
We try now to give a quantitative description of the possible scenario
introduced above. The goal is to find constraints on system parameters in 
order to match the measured luminosity and $\dot{\omega}_o/\omega_o$ and meet 
the requirement that the resulting regime has a sufficiently long lifetime.
%
\begin{figure}[h]
\includegraphics{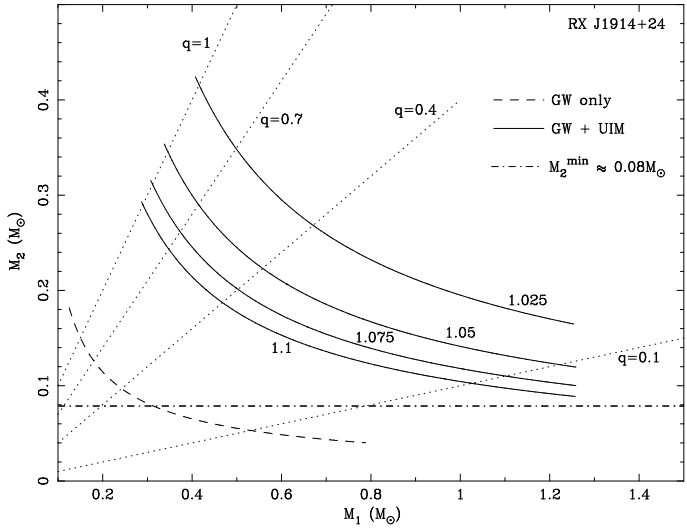}
\vspace{7.0cm}
\caption{$M_2$ vs. $M_1$ plot based on measured timing properties of
RX J1914+24. The dot-dashed line corresponds to the minimum mass for a 
degenerate secondary not to fill its Roche-lobe. The dashed curve represents 
the locus expected if orbital decay was driven by GW alone, with no spin-orbit.
coupling. It is seen that this curve is consistent with a 
detached system only if both masses are extremely low. The solid lines 
describe the loci expected if spin-orbit coupling for the primary 
star (the secondary spin is tidally locked to the orbit) is present and gives a
 \textit{negative} contribution to $\dot{\omega}_o$. The four curves are 
obtained assuming $W = 10^{33}$ erg s$^{-1}$ and four different values of 
$\alpha = 1.025, 1.05, 1.075$, $1.1$, respectively, from top to bottom, as 
reported in the plot.}
\label{fig5}
\end{figure}
Given that it is not possible to determine all system parameters uniquely, the 
scheme that we have used is as follows: given a value of $\alpha$, the measured
$\dot{\omega}_o, \omega_o$ and $W$ allow us to determine $\dot{\alpha}$ from 
eq. (\ref{alfaevolve}). For each value of M$_1$ eq. (\ref{omegadot}), 
re-written in the form reported in the Appendix, can thus be used to determine 
the value of M$_2$ (or $q$) that is compatible with $\dot{\omega}_o$ and 
$\omega_o$. This yields the solid curves of Fig. \ref{fig5}.\\
As these curves show, the larger is $\alpha$ and the smaller the upward shift 
of the corresponding locus, which may seem surprising. However, these curves 
are obtained at fixed luminosity $W$ and $\dot{\omega}_o$. Recalling 
that  $(1/\alpha)$ gives the efficiency of angular momentum transfer in systems
with $\alpha >1$ (cfr. $\S$\ref{efficiency}), a higher $\alpha$ at a given
luminosity implies that less energy and angular momentum are being transferred
to the orbit. Accordingly, GWs must be emitted at a smaller rate for a given 
$\dot{\omega}_o$.\\
The values of $\alpha$ in Fig. \ref{fig5} were chosen arbitrarily and are 
just illustrative: note that the resulting curves are quite similar to those
obtained for RX J0806+15.
Given $\alpha$, one can also estimate $k$ from the definiton of $W$ (eq. 
\ref{W}). On the other hand, the information given by the curves of Fig. 
\ref{fig5} yields determination of all quantities contained in $k$, apart from 
$\mu_1$.
Therefore, assuming $\overline{\sigma} = 3\times 10^{13}$ (e.s.u.) as in
the previous case, we can determine the value of $\mu_1$ along each of the
four curves of Fig. \ref{fig5}. Derived values are plotted in Fig. 
\ref{fig6}. We see that $\mu_1$ varies by just a factor 2-3 as a function of 
M$_1$ along each curve, while it depends somewhat more strongly on the assumed 
value of $\alpha$. \\
The resulting numbers correspond in any case to a moderately magnetic primary, 
indeed quite similar to what obtained for RX J0806+15 (the surface field B is 
in the $10^2 \div 10^5$ G range, depending on the primary mass).
%
\begin{figure}[h]
\includegraphics{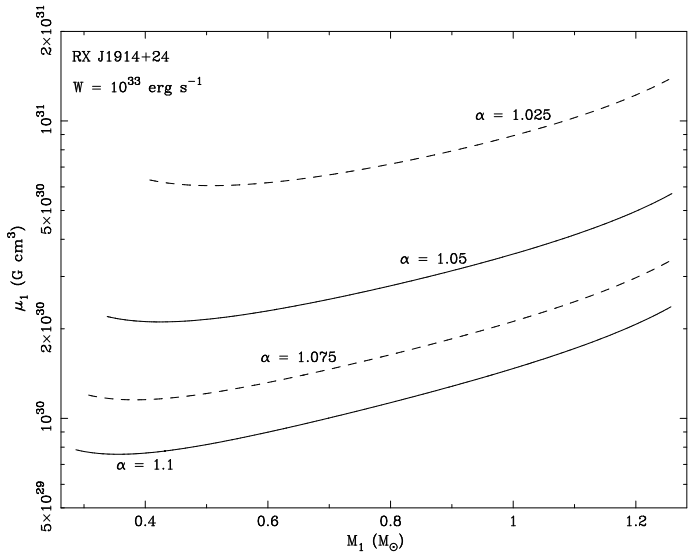}
\vspace{7.0cm}
\caption{$\mu_1$ as a function of the primary mass for 
the same values of $\alpha$ used previously and reported on the corresponding 
curves. Given a value of $\alpha$ and the estimated luminosity $W\sim 10^{33}$ 
erg s$^{-1}$, $\dot{\alpha}$ is calculated from eq. 
(\ref{alfaevolve}) as a function of M$_1$. $q$ as well is obtained as a 
function of M$_1$ from Fig. \ref{fig5}. All system parameters contained in $k$ 
are thus determined, apart from $\mu_1$. The estimated luminosity $W$ and the 
assumed value of $\alpha$ give the corresponding value of $k$: combining all 
the information, $\mu_1$ is self-consistently determined.}
\label{fig6}
\end{figure}
%
%
\subsection{Lifetime of the asynchronous state}
\label{lifetime}
\begin{figure}[h]
\includegraphics{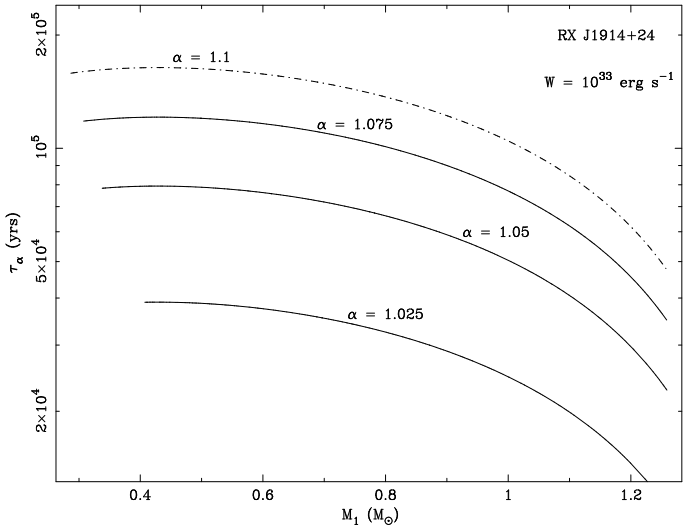}
\vspace{7.0cm}
\caption{The evolution timescale $\tau_{\alpha}$ as a function of the primary 
mass for the same values of $\alpha$ used previously, reported on the 
corresponding curves. Given the estimated luminosity $W\sim 10^{33}$ erg 
s$^{-1}$ and a value of $\alpha$, $\tau_{\alpha}$ is calculated from 
eq. (\ref{alfaevolve}) as a function of $I_1$ and, hence, M$_1$.} 
\label{fig8}
\end{figure}

Finally, having argued that this system can be explained in the UIM framework 
by a relatively highly asynchronous and moderately magnetic system, we have to 
deal with the timescale problem.\\
While RX J0806+15 is fully consistent with being in steady-state, so that it
has no timescale problem, we have demonstrated that RX J1914+24 cannot be in 
the same regime. Moreover, given the combination of a high luminosity and slow 
orbital evolution, we have found arguments in favour of its primary spinning 
faster than the orbital motion. It must be fast enough, indeed, that the 
orbital evolution is significantly slowed down by the angular momentum 
transferred to the orbit by spin-orbit coupling. Therefore, the system is in 
the fully non-linear regime where spin-orbit coupling is non negligible and the
evolution equation for $\omega_o$ must be solved explicitly.\\
We can estimate the synchronization timescale $\tau_{\alpha}$ starting from
the evolution equation (\ref{alfaevolve}). With the measured values of $W$, 
$\omega_o$ and $\dot{\omega}_o$, $\tau_{\alpha}$ can be calculated as a 
function of $I_1$ (and, thus, of M$_1$) given a particular value of $\alpha$.
In Fig. \ref{fig8} we have reported the results obtained for the same four
values of $\alpha$ assumed in the previous section. \\
It is apparent that the larger $\alpha$, the longer the timescale, for a given 
value of M$_1$. This is because the system luminosity is a fixed (measured)
quantity: therefore the larger the asynchronism, the smaller the efficiency of 
spin-orbit coupling and, accordingly, the larger $\tau_{\alpha}$.
%
%
The resulting timescales range from a few $\times 10^4$ yrs to a few $\times 
10^5$ yrs, while $\tau_o \sim 6\times 10^6$ yrs for this source. Although 
$\tau_{\alpha}$ may still seem too short, we stress that in this scenario the 
long orbital evolutionary timescale results largely from the effect of 
spin-orbit coupling. The latter decreases strongly on the timescale 
$\tau_{\alpha}$, after which $\tau_o$ is essentially determined by GW emission 
only. We denote the latter as $\tau^{GW}_o$. Therefore, the 
system will actually spend the total time $\tau^{GW}_o$ at the nearly constant 
orbital period, its evolution being slowed down by just a fraction 
$\tau_{\alpha}/ \tau^{GW}_o$ of this time.
\begin{figure}[h]
\includegraphics{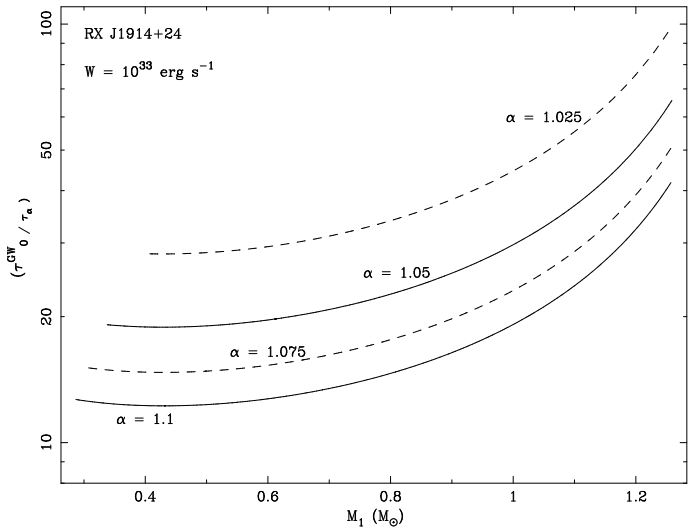}
\vspace{7.0cm}
\caption{The ratio $\tau^{GW}_o / \tau_{\alpha}$ calculated for the four
values of $\alpha = 1.025, 1.05, 1.075, 1.1$. $\tau^{GW}_o$ is obtained from 
the value of $X$ along each of the four curves of Fig. \ref{fig5}, according to
eq. (\ref{taugw}), while the corresponding value of $\tau_{\alpha}$ are 
those plotted in Fig. \ref{fig8}.} 
\label{fig9}
\end{figure}
In Fig. \ref{fig9} we have plotted the ratio $(\tau^{GW}_o/ \tau_{\alpha})$ 
for the four cases studied above (Figs. \ref{fig5}, \ref{fig6}
and \ref{fig8}). The values $\tau^{GW}_o$ were calculated along the solid 
curves of Fig. \ref{fig5}, that define the value of $X$ for each (M$_1$,M$_2$).
Given $X$, both the source GW luminosity $\dot{E}_{\mbox{\tiny{gr}}}$ and 
$\tau^{GW}_o$ can be calculated. In particular, since $g(\omega_o)$ is a 
function of $X$ itself, we have
\begin{equation}
\label{taugw}
\left(\frac{\dot{\omega}_o}{\omega_o}\right)^{GW} = (\tau^{GW}_o)^{-1} =
\frac{96}{5} \frac{G^{5/3}}{\mbox{c}^5} \omega^{8/3}_o X~.
\end{equation}
According to the above reasoning, the expected GW luminosity of this source is 
in the range $(4.6 \div 1.4) \times 10^{34}$ erg s$^{-1}$. The corresponding 
ratios $\dot{E}_{\mbox{\tiny{gr}}}/ \dot{E}^{(orb)}_L$ are $1.15, 1.21, 1.29$
and $1.4$, respectively, for $\alpha =1.025, 1.05, 1.075$ and $1.1$.\\
Therefore, RX J1914+24 and RX J0806+15 may even have similar component masses 
and magnetic moments. The significant observational differences would 
essentially be due to the two systems being caught in different evolutionary 
stages. RX J1914+24 would be in a luminous, transient phase that preceeds its 
settling into the dimmer steady-state, a regime already reached by the shorter 
period RX J0806+15.
Although the more luminous phase is transient, its lifetime can be as long as 
$ \sim 10^5$ yrs, one or two orders of magnitude longer than previously 
estimated.\\
%
%
We comment here on the two symmetric peaks, around the main one, found 
in a few power spectra by Ramsay et al. (2006).\\
In the UIM framework, given a primary spin $\omega_1$ that differs from 
$\omega_o$, some signature of both periodicities may be expected, indeed. The 
hot spot position on the primary surface is modulated at the orbital period 
(the spot always follows the position of the secondary star) and at the beat 
period $\omega_b =\omega_1 - \omega_o$. The latter because the spot position 
is determined by the position of the primary's magnetic pole as well, which 
itself rotates at $\omega_1$. If the field simmetry axis were perfectly aligned
to the spin (and orbital) axis, this second modulation would disappear but, in 
general, some inclination is to be expected. Barros et al. (2004) have shown 
that, in the context of the UIM, the modulation at $\omega_b$ in RX J1914+24 is
small, possibly negligible, if the inclination angle $\beta < (15-19)
^{\mbox{\tiny{o}}}$. However, it would be a natural expectation of the model to
have two main periodicities in power spectra, namely $\omega_o$ and $\omega_b$:
when $\omega_1 \sim \omega_o$ (thus $\alpha \sim 1$) the frequency $\omega_b 
\ll \omega_o$.\\
This is highly speculative at this stage and proper quantitative studies of 
this phenomenon, both on the observational and theoretical side, are necessary 
before these features can be used to put quantitative constraints on models. 
These are beyond the scope of this paper and are currently in progress. What we
want to stress here is that a second periodicity in power spectra is not a 
finding against the UIM. The features reported by Ramsay et al. (2006) are not 
in constrast with qualitative expectations in the UIM.
\section{Dependence of the results on the assumed luminosity}
\label{change}
In this section we briefly discuss how our results change when different source
luminosities are assumed, reflecting the large uncertainties in their distances
and bolometric emission. We have repeated the same reasoning presented in the 
previous sections, simply changing the adopted value of W to a significantly 
larger value.
\subsection{RX J0806+15}
We have discussed in $\S$\ref{rxj08} the uncertainty in the bolometric 
luminosity of this source, which could possibly be up to a factor 5-6 larger 
than the X-ray emission reported by Israel et al (2003). We assume here 
W =$ 5 \times 10^{32}$ erg s$^{-1}$ and see how our conclusions change.\\
For this luminosity, a six times larger degree of asynchronism than previously
assumed would result, $1-\alpha_{\infty} \simeq 3.8 \times 10^{-2}$, still
a quite small number. Given this value, the corresponding locus in the M$_2$ 
vs. M$_{1}$ plane would correspond to values of $X$ and, accordingly, GW 
luminosities smaller by just a few percent. The corresponding primary magnetic 
moment and implied surface field are themselves decreased by the same factor 
of six by which $W$ increases. All these changes, although non-negligible on 
quantitative grounds, do not affect significantly the overall picture presented
above.
\subsection{RX J1914+24}
The case of RX J1914+24 is more difficult: the luminosity assumed in 
$\S$ \ref{rxj19} lies at the centre of a very wide range of proposed values, 
from the smallest W $\simeq 3 \times 10^{31}$ erg s$^{-1}$ (Ramsay et al. 2006)
to the largest W $\simeq 10^{35}$ erg s$^{-1}$ (Ramsay et al. 2005).
\begin{figure}[h]
\includegraphics{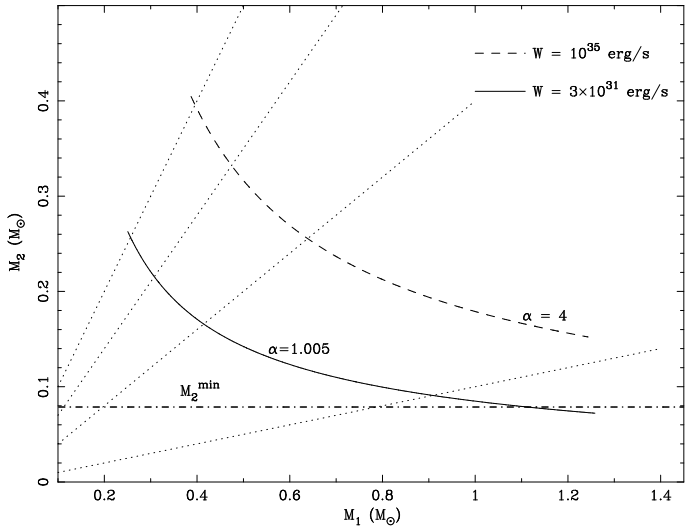}
\vspace{7.0cm}
\caption{The loci in the M$_2$ vs. M$_1$ plane obtained assuming a luminosity 
of $W=3\times 10^{31}$ erg/s and $\alpha =1.005$ (solid line) or a very high
luminosity of $\sim 10^{35}$ erg/s and $\alpha =4$ (dashed line). As noted in
the previous section, curves shift upwards for smaller values of $\alpha$ and
downwards for increasing values of the asynchronism.}
\label{locichange}
\end{figure}
We show and briefly discuss here the same figures presented in the previous 
section, obtained for just the two extreme luminosities above. Instead of 
exploring a range of values of $\alpha$, we are exploring the range of values 
of $W$, over approximately four orders of magnitude. Therefore just one curve,
is reported for each of the two cases. In all 
figures below the solid lines correspond to the low luminosity W$ = 3\times 
10^{31}$ erg s$^{-1}$, while the dashed curves represent the opposite case with
W$= 10^{35}$ erg s$^{-1}$.
\begin{figure}[h]
\includegraphics{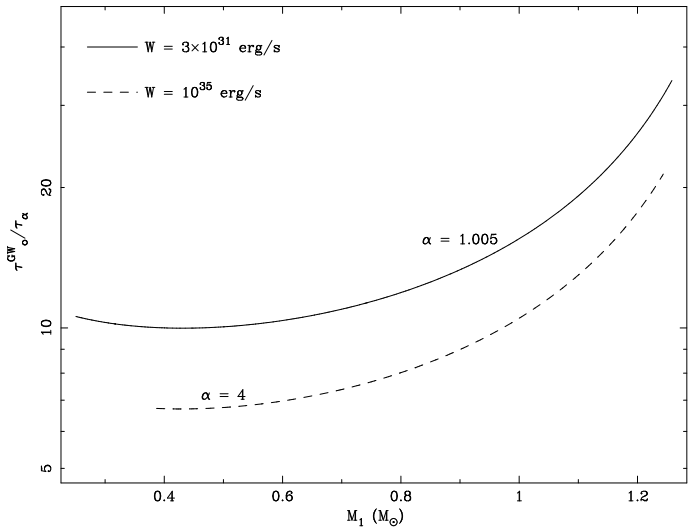}
\vspace{7.0cm}
\caption{The ratio $\tau^{GW}_o / \tau_{\alpha}$ calculated for W$= 3\times 
10^{31}$ erg/s and $\alpha = 1.005$ (solid line) and for W$=10^{35}$ erg/s and 
$\alpha=4$ (dashed line). $\tau^{GW}_o$ is obtained from the value of $X$ along
the corresponding curve of Fig. \ref{locichange}, according to eq. 
(\ref{taugw}), while the corresponding value of $\tau_{\alpha}$ are 
calculated as described in the caption to Fig. \ref{fig8}.} 
\label{raziochange}
\end{figure}
From Fig. \ref{locichange} we see that the high luminosity case would require
a significant degree of asynchronism, but  still of the order of a few, not the
$\gg 100$ value previously suggested in the literature. On the other hand, the
very dim case can be well described by a low asynchronism, even lower than
the steady-state asynchronism found for RX J0806+15. However, we have shown 
that $1-\alpha_{\infty}\simeq 0.15$ would be implied for this source if
$3 \times 10^{31}$ erg s$^{-1} =$ W$_{\infty}$. Rather, $\alpha=1.005$ gives 
$(1-\alpha_{\infty}) \simeq 1.7 \times 10^{-4}$ (eq. \ref{j19}).\\ 
The dimmer system would have a long lifetime, $\tau_{\alpha} \sim 3\times 10^5$
yrs, which is $\sim$ one tenth to one twentieth of the orbital evolutionary 
timescale $\tau^{GW}_o$. From the solid curve of Fig. \ref{locichange} we can 
estimate a GW luminosity $\dot{E}_{\mbox{\tiny{gr}}} \simeq 10^{34}$ erg 
s$^{-1}$ and a ratio $\dot{E}_{\mbox{\tiny{gr}}} / \dot{E}^{(orb)}_L \sim 1.6$.
\begin{figure}[h]
\includegraphics{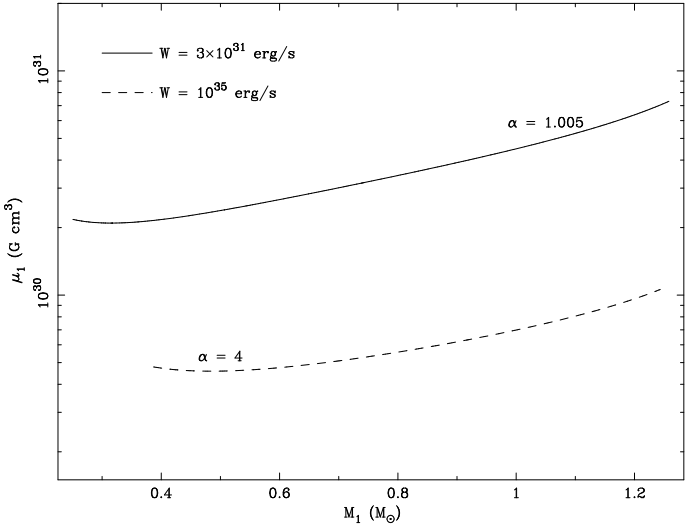}
\vspace{7.0cm}
\caption{The expected value of the primar's magnetic moment $\mu_1$ as a 
function of the primary mass (M$_1$) for the same two extreme luminosities of 
the above figures, indicated in the figure. Despite the non negligible
difference between the two curves, these still span the range of medium-to-weak
magnetization already found previously.}
\label{muchange}
\end{figure}
On the other hand, the very luminous system has a shorter 
$\tau_{\alpha} \sim$ a few $\times 10^4$ yrs, but can still have a lifetime 
$\sim$ one tenth of the orbital evolutionary timescale $\tau^{GW}_o$. The 
latter is itself significantly shorter than $\omega_o/\dot{\omega}_o$, because 
large component masses would be required and, correspondingly, very strong GW 
emission expected. For the case with $\alpha =4$ we estimate 
$\dot{E}_{\mbox{\tiny{gr}}} \simeq 3.6 \times10^{34}$ erg s$^{-1}$, with a 
ratio $\sim 1.25$ to the rate of work $W/(1-\alpha)$.\\
Finally, Fig. \ref{muchange} shows the expected primary magnetic moments in 
both cases. 
\section{Conclusions}
\label{conclusions}
In the present work we have clarified some key aspects of spin-orbit coupling
in the Unipolar Inductor Model coupled to GW and applied the model to the two 
candidate ultrashort period binaries RX J0806+15 and RX J1914+24. 
We derive constraints
%
%
on system parameters in order for the model to account for the inferred 
luminosities and measured $\omega_o$ and $\dot{\omega}_o$ of both sources.
The UIM accounts in a natural way for their luminosities, measured orbital
periods and rates of orbital spin-up, provided they have moderately 
magnetic primaries, say $\mu_1 \sim~\mbox{several}\times 10^{29}~\mbox{to few} 
\times 10^{30}$ G cm$^3$. Concerning each of the sources, we conclude that:
\begin{itemize}
\item
In the case of RX J0806+15 the observations are explained in the framework of
the steady-state solution. The degree of asynchronism is determined by a 
balance between the rate of current dissipation and the rate at which energy is
fed to the electrical circuit by orbital evolution, while the source luminosity
does not decrease at all. The orbital evolution is driven mostly by GW-emission
and the system's timing properties should differ only slightly from those of 
two point masses evolving under GW emission. The modest primary magnetic moment
represents a key requirement of this interpretation and its measurement would 
be a crucial test as to its viability.

\item RX J1914+24, on the other hand, cannot have reached the steady-state
solution yet, as its present rate of electric energy dissipation is 
significantly higher than what GW emission alone could feed to the circuit. \\
In addition to requiring a moderately magnetic primary (again $\mu_1 \sim 
10^{30}$ G cm$^3$), we conclude that the primary spin frequency must be 
somewhat larger than the orbital frequency. For W$\simeq 10^{33}$ erg s$^{-1}$ 
the parameter $(1-\alpha)$ can be in the range $(0.1 \div 0.02)$ (P$_{\mbox
{\tiny{spin}}}$ between 517 and 558 s). Although the source is not 
consistent with being in a steady-state we have shown that, for $\alpha$ in the
above range, the asynchronous, luminous and transient phase can last between 
a few $\times 10^4$ yrs up to $3\times 10^5$ yrs. These are between 2\% and 
10\% of the orbital evolution timescale. Therefore, there is a sizeable 
probability that one in two of the ultrashort period binaries that have so far 
been found is in this evolutionary stage.\\
The measured value of $\dot{\omega}_o$ does not reflect the actual rate of GW 
emission from this system. It rather results from the sum of GWs and spin-orbit
coupling, the latter partially cancelling the effect of the former. For the 
range of values of $\alpha$ that we consider more likely, $\alpha \simeq 
(1.025\div 1.1$), the actual GW luminosity should be in the $\sim 10^{34}$ erg 
s$^{-1}$ range. \\
Values of $\alpha$ between 1.005 and $\sim 4\div 5$ can explain all possible 
source luminosities in the range ($3\times 10^{31} \div 10^{35}$) erg s$^{-1}$.
In particular, $\alpha \sim$ of a few can self-consistently explain a 
luminosity as high as $10^{35}$ erg s$^{-1}$, extremely less then previously 
claimed in the literature.
\end{itemize}
In this work we were mainly concerned with finding system parameters for which 
the model could be made consistent with the energetic requirements and timing 
properties of the candidate sources. Further work is needed to address issues 
such as the duration and shape of the X-ray pulsations. 
Predictions that can be checked through observations to be carried out in the 
next few years are essentially of three kinds:
\begin{itemize}
\item independent measurements of the primary magnetic moment $\mu_1$, which 
should result to be smaller than several $\times 10^{30}$ G cm$^3$ in order for
the model to account for the two sources. Although several details of the model
should still be taken into account, it seems difficult that the overall 
energetic requirements could be changed by orders of magnitude. The expectation
of low magnetic moments seems therefore quite robust.

\item extending the temporal baseline over which timing studies are carried out
may help detecting the second derivative of $\omega_o$ in both systems, but 
particularly in RX J1914+24. Such measurements could provide an indipendent 
estimate of system masses and/or confirm a significant contribution of UI to 
the orbital evolution.

\item a detailed study of the duration and shape of the X-ray pulsations, as
well as of power spectra, would be required, in order to put constraints on 
geometric properties of the systems, possibly related to the shape and 
orientation of magnetic field lines near the primary component.\\
Account for hot spot shapes different from a simple filled spot should be taken
since two detached, arc-like hot regions are expected instead (see Wu et al. 
2002).
\end{itemize}
Concerning the longer term future, we emphasize that LISA is expected to 
measure the GW emission of very short period binaries such as those discussed
here, therefore yielding an independent check of the model. In particular,
RX J1914+24 is expected to be much more luminous in GWs than one would infer 
based on its measured $\omega_o$ and $\dot{\omega}_o$.
\section{Appendix}
\label{appendice}
We derive here the expression used to obtain the solid curves of Fig. 
\ref{fig5}. In the case of Fig. \ref{fig3}, $\alpha$ was assumed to be 
constant, as this describes the relevant condition to RX J0806+15. 
On the other hand, in studying RX J1914+24 we were led to deal with the more 
general case in which $\alpha$ has a non negligibile time derivative.\\
Let $g(\omega_o)$ be defined as:
\begin{equation}
\label{g}
g(\omega_o) = -\frac{1}{3} \left[\frac{q^3 G^2 M^5_1 \omega^{2}_o}{1+q}
\right]^{1/3}\left[1- \frac{6}{5} (1+q) \left(\frac{R_2}{a}\right)^2\right]
\end{equation}
where the second term in square brackets is denoted by $f(\omega_o)$.
Then, from eq. (\ref{omegadot}), (\ref{alfaevolve}) and the definition of
$g(\omega_o)$:
\begin{equation}
\label{appendice1}
\frac{\dot{\omega}_o}{\omega_o} = \frac{1}{g(\omega_o)} \left[
\dot{E}_{gr} - \left(\frac{\dot{\alpha}}{\alpha} + \frac{\dot{\omega}_o}
{\omega_o}\right) \alpha I_1 \omega^2_o\right]
\end{equation}
and from this, re-writing the pure-GW period derivative as $(\dot{\omega}_o / 
\omega_o)_g = \dot{E}_{gr}f(\omega_o) / g(\omega_o) $:
\begin{equation}
\label{appendice2}
\frac{\dot{\omega}_o}{\omega_o} \left[1+ \alpha  
\frac{I_1 \omega^2_o}{g(\omega_o)}\right]  = \left(\frac{\dot{\omega}_o}
{\omega_o}\right)_{g} \left[\frac{1}{f(\omega_o)} - \frac{\dot{\alpha}I_1 
\omega^2_o}{g(\omega_o) \left(\frac{\dot{\omega}_o} {\omega_o}\right)_g}
\right]
\end{equation}
that leads to the following expression:
\begin{equation}
\label{appendice3}
\frac{\dot{\omega}_o}{\omega_o} =  \frac{\left(\dot{\omega}_o / 
\omega_o \right)_g} {g(\omega_o) + \alpha I_1 \omega^2_o} 
\left[\frac{g(\omega_o)}{f(\omega_o)} -\dot{\alpha} \frac{I_1 \omega^2_o}
{(\dot{\omega}_o / \omega_o)_g}\right]
\end{equation}
Now substitute the expressions for $g(\omega_o)$ and $f(\omega_o)$, 
remembering that $I_1 = (2/5) M_1 R^2_1$ and that, from Kepler's third law, 
$\omega^{-8/3}_o = [GM_1 (1+q)]^{-4/3} a^4$. After some algebra it obtains:
%
\begin{eqnarray}
\label{appendice4}
\frac{\dot{\omega}_o}{\omega_o} & = &\left(\frac{\dot{\omega}_o}{\omega_o}
\right)_g \left[1+ \frac{\dot{\alpha}}{16 q^2} \left(\frac{R_1}{a}\right)^2 
\frac{c^5 a^4}{(GM_1)^3}\right] \nonumber \\ 
 & & \times \left[1- \frac{6}{5} \alpha\frac{1+q}{q} \left(\frac{R_1}{a}
\right)^2 - \frac{6}{5} (1+q) \left(\frac{R_2}{a}\right)^2\right]^{-1}
\end{eqnarray}
The solid curves of Fig. \ref{fig5} were obtained just using eq. 
(\ref{appendice4}). This takes account of the primary asynchronism and its 
rate of variation as well, while assumes the secondary to be perfectly 
synchronous (tidally locked) at all times.

\end{document}